\documentclass[12pt,english,draftcls, one column]{IEEEtran}
\usepackage[T1]{fontenc}
\usepackage[latin9]{inputenc}
\usepackage{float}
\usepackage{amsmath}
\usepackage{amsthm}
\usepackage{amssymb}
\usepackage{graphicx}
\usepackage{stfloats}
\usepackage{cite} 
\usepackage{xcolor}

\usepackage{cite}

\makeatletter

\floatstyle{ruled}
\newfloat{algorithm}{tbp}{loa}
\providecommand{\algorithmname}{Algorithm}
\floatname{algorithm}{\protect\algorithmname}

\theoremstyle{plain}

\theoremstyle{plain}

\ifCLASSINFOpdf
\else
\fi
\usepackage{babel}
\usepackage{graphicx}
\usepackage{ulem}
\usepackage{xcolor}

\makeatother

\usepackage{babel}
\providecommand{\propositionname}{Proposition}
\providecommand{\theoremname}{Theorem}

\begin{document}

\title{Completion Time Minimization of\\ Fog-RAN-Assisted Federated Learning With Rate-Splitting Transmission}

\author{Seok-Hwan Park, \textit{Member}, \textit{IEEE}, and Hoon Lee, \textit{Member}, \textit{IEEE} \thanks{
This work is supported in part by Institute of Information $\&$ communications Technology Planning $\&$ Evaluation (IITP) grant funded by the MSIT (No. 2021-0-00467, Intelligent 6G Wireless Access System); in part by the National Research Foundation (NRF) of Korea Grants funded by the MOE under Grant 2019R1A6A1A09031717 and by the MSIT under Grant 2021R1C1C1006557. \textit{(Corresponding author: Hoon Lee.)}

Copyright (c) 2015 IEEE. Personal use of this material is permitted. However, permission to use this material for any other purposes must be obtained from the IEEE by sending a request to pubs-permissions@ieee.org.

S.-H. Park is with the Division of Electronic Engineering, Jeonbuk
National University, Jeonju, Korea (email:  seokhwan@jbnu.ac.kr).

H. Lee is with the Department of Information and Communications Engineering, Pukyong National University, Busan 48513, South Korea (e-mail: hlee@pknu.ac.kr).}}
\maketitle

\begin{abstract}
This work studies federated learning (FL) over a fog radio access network, in which multiple internet-of-things (IoT) devices cooperatively learn a shared machine learning model by communicating with a cloud server (CS) through distributed access points (APs). Under the assumption that the fronthaul links connecting APs to CS have finite capacity, a rate-splitting transmission at IoT devices (IDs) is proposed which enables hybrid edge and cloud decoding of split uplink messages. The problem of completion time minimization for FL is tackled by optimizing the rate-splitting transmission and fronthaul quantization strategies along with training hyperparameters such as precision and iteration numbers. Numerical results show that the proposed rate-splitting transmission achieves notable gains over benchmark schemes which rely solely on edge or cloud decoding.
\end{abstract}

\begin{IEEEkeywords}
Federated learning, Fog-RAN, rate splitting, hybrid decoding, completion time minimization.
\end{IEEEkeywords}


\theoremstyle{theorem}
\newtheorem{theorem}{Theorem}
\theoremstyle{proposition}
\newtheorem{proposition}{Proposition}
\theoremstyle{lemma}
\newtheorem{lemma}{Lemma}
\theoremstyle{corollary}
\newtheorem{corollary}{Corollary}
\theoremstyle{definition}
\newtheorem{definition}{Definition}
\theoremstyle{remark}
\newtheorem{remark}{Remark}

\section{Introduction}

To alleviate the constraints on privacy and communication resources, centralized machine learning algorithms are envisioned  to be replaced with distributed learning frameworks, in which distributed internet-of-things (IoT) devices collaboratively train a shared machine learning model while communicating with a cloud server (CS) \cite{Park:Proc21}.
In federated learning (FL), which is one of the most emerging distributed learning frameworks, IoT devices (IDs) upload locally trained models to CS \cite{Konecny:arXiv16, Vu:TWC20, Yang:TWC21, Al-Abiad:arXiv21, Yao:TCCN}. Accordingly, we can protect the privacy of local dataset and save the spectrum resource.
To avoid long latency required for exchange of local and global model updates between IDs and CS, edge learning has been actively studied, which enjoys low latency by bringing the decision-making tasks closer to IDs \cite{Park:Proc19}. However, proximate edge servers should perform partial model aggregation with limited computing powers.
In this work, we focus on the cloud-based FL at CS, which cooperates with a wider range of IDs and has more access to data.
To alleviate the communication bottleneck, we address the signal processing design of IDs, distributed access points (APs), and CS for low-latency transmission on wireless and fronthaul links.

In \cite{Yang:TWC21}, wireless communication for FL was studied under a single-cell scenario, in which IDs communicate with a single AP directly connected to a CS.
More complicated scenarios with multiple APs were discussed in \cite{Vu:TWC20, Al-Abiad:arXiv21}.
The problems of minimizing the energy consumption at IDs were tackled in \cite{Yang:TWC21, Vu:TWC20}, while the completion time minimization for FL was addressed in \cite{Al-Abiad:arXiv21}. 
For the transmission over wireless channel, frequency division multiple access (FDMA) was assumed in \cite{Yang:TWC21}, and non-orthogonal transmission across IDs was considered in \cite{Vu:TWC20, Al-Abiad:arXiv21}.
Regarding the fronthauling strategy, \cite{Vu:TWC20} assumed that the fronthaul links can carry the uplink received signals of APs to the CS without distortion and latency.
This assumption can be made possible only when the fronthaul links have sufficiently large capacity compared to the uplink bandwidth.
The work in \cite{Al-Abiad:arXiv21} addressed a more challenging and realistic case, where the capacity of fronthaul links is limited. 
The edge decoding strategy is considered, whereby each AP decodes the messages sent by the associated IDs and transmits the decoded information on fronthaul. 
However, when the fronthaul links have sufficient capacity, cloud decoding would be a more efficient strategy, in which the fronthaul links carry quantized and compressed version of the uplink signals, and a joint cloud decoding at CS is followed.

In this work, we study FL over a fog radio access network (RAN), in which multiple IDs cooperate to handle a machine learning task by repeatedly uploading locally updated models to a CS through distributed APs.
We aim at achieving the advantages of both the edge and cloud decoding strategies by adopting rate-splitting transmission at IDs. 
The effectiveness of the rate-splitting approach has been reported in \cite{Hao:TCOM17, Mao:EURASIP18, Mao:TCOM21, Mao:arXiv22} in terms of overcoming the impairments caused by channel state information (CSI) uncertainty and generalizing existing multiple access techniques. 
The degrees-of-freedom (DoFs) region of the rate-splitting scheme was analyzed in \cite{Hao:TCOM17} for multi-cell multiple-input single-output (MISO) interference channel (IC) with arbitrary CSI quality.
A generalized linearly precoded multiple access scheme referred to as rate-splitting multiple access was proposed in \cite{Mao:EURASIP18} which subsumes conventional space-division and non-orthogonal multiple access schemes as special cases.
In \cite{Mao:TCOM21}, the advantages of the rate-splitting transmission were studied for MISO broadcast channel (BC) with heterogeneous CSI qualities among users in terms of DoFs and average sum-rate. 
We refer to \cite{Mao:arXiv22} for a comprehensive overview of rate-splitting techniques including appealing applications and standardization issues.

In this work, we study an application of the rate-splitting transmission for fog-RAN-assisted FL.
In the proposed rate-splitting scheme, each ID splits its message into two submessages, which are encoded to independent data streams, and the two encoded signals are delivered to CS by means of superposition coding and edge and cloud decoding strategies.
We minimize the competition time of the FL algorithm. Our design includes the joint optimization of the rate-splitting transmission and fronthaul quantization strategies along with hyperparameters of training algorithms, in particular, the algorithm precision and iteration numbers satisfying a target learning performance.
Numerical results validate the advantages of the proposed FL system in terms of completion time.


\section{Federated Learning System Over Fog-RAN\label{sec:System-Model}}

We consider an FL system in a fog network, in which a CS and $N_I$ IDs collaboratively learn a machine learning model through $N_A$ APs. The IDs and APs, each equipped with $M_{I}$ and $M_{A}$ antennas, communicate on wireless access channels. The APs are connected to the CS via orthogonal digital fronthaul links each of capacity $C_F$ bits per second (bps).
We define the index sets $\mathcal{N}_I\triangleq\{1,\ldots,N_I\}$ and $\mathcal{N}_A\triangleq\{1,\ldots,N_A\}$. The goal of FL is to find a parameter vector $\mathbf{w}\in\mathbb{R}^{d}$ of a machine learning model with distributed datasets at IDs. Such a problem can be formulated as the minimization of a global loss function $f_L(\mathbf{w}) = (1/\sum_{l\in\mathcal{N}_I}D_l) \sum_{k\in\mathcal{N}_I} D_k f_{L,k}(\mathbf{w})$ where $f_{L,k}(\mathbf{w})$ represents the local loss function computed at ID $k$ using $D_{k}$ data samples.

\begin{figure}
\centering\includegraphics[width=0.8\linewidth]{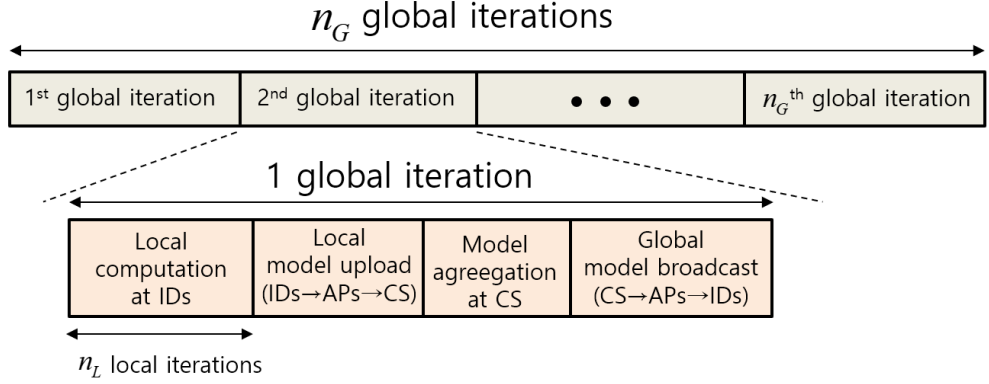}
\caption{{\label{fig:overall-procedure}The overall procedure of the fog-RAN-assisted FL system.}}
\end{figure}

Fig. \ref{fig:overall-procedure} illustrates the overall procedure of the FL containing $n_{G}$ global iterations. Each global iteration consists of four steps:
local computation at IDs; local model upload from IDs to CS through APs; model aggregation at CS; and global model broadcast from CS to IDs through APs.
Individual IDs update their local parameter vectors using, e.g., the gradient descent (GD) method, with $n_{L}$ local iterations. Updated models are then sent to distributed APs through wireless uplink channels, and each AP conveys received local models to the CS via capacity-constrained fronthaul links. After aggregating local models, the CS broadcasts a global model to all IDs. These are repeated for $n_{G}$ global iterations.

The iteration numbers $n_{G}$ and $n_{L}$ depend on the target accuracy levels and the adopted FL algorithm. 
According to the analysis in \cite{Yang:TWC21}, these are lower-bounded as
\begin{align}
    n_{L} &\geq \frac{2}{ (2 - \beta\delta) \delta \alpha} \log_2 \frac{1}{\eta_L}, \label{eq:number-local-iterations} \\
    n_{G} &\geq  \frac{2 \beta^2 }{ \alpha^2 \xi (1 - \eta_L) } \ln \frac{1}{\eta_G}, \label{eq:number-global-iterations}
\end{align}
where $\eta_L$ and $\eta_G$ respectively indicate the target accuracy levels for local and global optimization problems, and $\delta$ denotes the learning rate. Constants $\alpha$, $\beta$, and $\xi \in (0,\alpha/\beta ]$ are determined by the properties of the local training problem.
Following \cite{Yang:TWC21, Yao:TCCN}, we set $\alpha=2$, $\beta=4$, $\delta=1/4$, and $\xi=1/10$, but the optimization algorithm proposed in this work can be applied for arbitrary values. 
The global accuracy target $\eta_{G}$ is related to the number of the global FL iterations $n_{G}$, and thus it can be fixed in advance. On the contrary, the local accuracy $\eta_{L}$ affects both the global and local iterations and invokes a nontrivial tradeoff. Reducing the local accuracy target $\eta_{L}$, which imposes a smaller deviation from optimal solutions, requires a more local iterations $n_{L}$, but at the same time, can reduce the global FL repetitions $n_{G}$. Thus, we need to carefully design $n_{L}$ along with the transmission strategy over wireless and fronthaul access links.

Thanks to the powerful computing resources of CS and higher transmission power of APs, the latency of the model aggregation and global model broadcast steps is negligible compared to that of the operations of IDs \cite{Yang:TWC21, Al-Abiad:arXiv21, Yao:TCCN}. Thus,
we can model the completion time $\tau_{\text{total}}$ as
\begin{align}
    \tau_{\text{total}} = n_{G} \left( \tau_C + \tau_{W} + \tau_{F} \right), \label{eq:total-latency}
\end{align}
where $\tau_C$ is the local computation time and $\tau_{W}$ and $\tau_{F}$ respectively denote the latency associated with wireless and fronthaul access of the local model upload step.
Let $c_k$ be the local computing capacity of ID $k$ in CPU cycles per second. It is assumed that $n_C$ CPU cycles are required for processing one data sample.
Then, the local computation time at ID $k$ is given as $n_{L} n_C D_k / c_k$. The overall computation time $\tau_C$ is thus dominated by the worst-case ID as
\begin{align}
    \tau_{C} = \max_{k\in\mathcal{N}_{I}} \, \frac{n_{L} n_C D_k}{ c_k }. \label{eq:local-computing-time}
\end{align}

We express the received signal vector $\mathbf{y}_i\in\mathbb{C}^{{M_A}}$ of AP $i$ on the wireless uplink channel of bandwidth $B$ Hz as
\begin{align}
    \mathbf{y}_i = \sum\nolimits_{k\in\mathcal{N}_I} \mathbf{H}_{i,k} \mathbf{x}_k + \mathbf{z}_i, \label{eq:received-signal-AP-i}
\end{align}
where $\mathbf{H}_{i,k}\in\mathbb{C}^{{M_A\times M_I}}$ denotes the channel matrix from ID $k$ to AP $i$, $\mathbf{x}_k\in\mathbb{C}^{{M_I}}$ represents the transmitted signal of ID $k$ encoding its local model parameter, and $\mathbf{z}_i\sim\mathcal{CN}(\mathbf{0}, \sigma_z^2\mathbf{I})$ is the noise at AP $i$.
All IDs are subject to the transmit power constraint $\mathbb{E} \left[ ||\mathbf{x}_k||^2 \right] \leq P_{\text{tx}}$. We assign a serving AP $i_k\in\mathcal{N}_A$ for ID $k$ to process the signal $\mathbf{x}_{k}$. For given $i_1, i_2, \ldots, i_{N_I}$, we define the set $\mathcal{N}_{I,i} \subseteq \mathcal{N}_I$ of IDs served by AP $i$, i.e.,
\begin{align}
    \mathcal{N}_{I,i} = \{ k\in\mathcal{N}_I | i_k = i \}. \label{eq:set-ID-served-by-AP-i}
\end{align}
Accordingly, the edge-decoding signals $\mathbf{s}_{E,k}$ of the IDs $k\in\mathcal{N}_{I,i}$ are decoded by EN $i$ and forwarded to the CS.
In this work, we assume that the the AP-ID association $\mathcal{N}_{I,i}$ is arbitrarily predetermined and leave the optimization of those variables for future work.

\section{Rate-Splitting Transmission of Local Models} \label{sec:rate-splitting-transmission}

This section discusses the process of transmitting the locally updated models at IDs to the CS through distributed APs. One straightforward approach is a hard selection of decoding nodes for individual IDs. Each ID $k$ encodes the parameter vector of its local model into a single data message that can be decoded either by the serving AP $i_k$ (i.e., edge decoding) or the CS (i.e., cloud decoding). 
If the capacity $C_F$ of fronthaul links is sufficiently large, it is desirable to select cloud decoding whereby the CS utilizes the uplink received signals of all the connected APs for centralized interference management.
However, if the fronthaul capacity $C_F$ is limited, the uplink signals reported to CS via fronthaul may contain quantization distortion noise of significant power. For this reason, for small $C_F$, it is better to perform local edge decoding at APs avoiding the quantization distortion, even at the cost of the degradation by interference signals \cite{Stephen:TVT17}.
This observation invokes the combinatorial optimization of selecting one of the edge decoding and cloud decoding strategies for every ID, requiring an exhaustive search over $2^{N_{I}}$ possible cases.

To achieve the advantages of both decoding strategies avoiding a complicated search, we adopt the rate-splitting transmission \cite{Hao:TCOM17, Mao:TCOM21} which softly combines these two extreme decoding policies. Such a hybrid design facilitates an optimized APs-CS coordination for the successful decoding of local machine learning models, thereby improving the quality of the model aggregation at the CS. This approach is particularly beneficial for finite $C_{F}$ that suffer from the nontrivial tradeoff between fronthaul overhead and centralized interference management. The local model of ID $k$ is assumed to be represented in $d_k$ bits. Unlike the straightforward approach described above, ID $k$ divides the $d_k$ bits into two bit strings of size $d_{E,k}$ and $d_{C,k}$ bits, which satisfy
\begin{align} 
    d_{E,k} + d_{C,k} = d_k, \,\, d_{E,k} \in [0,d_k]. \label{eq:bit-split-conditions}
\end{align}
These bit strings are separately encoded to data signals $\mathbf{s}_{E,k}\in\mathbb{C}^{{M_{I}}}$ and $\mathbf{s}_{C,k}\in\mathbb{C}^{{M_{I}}}$ which are distributed as $\mathbf{s}_{E,k}\sim\mathcal{CN}(\mathbf{0}, \mathbf{Q}_{E,k})$ and $\mathbf{s}_{C,k}\sim\mathcal{CN}(\mathbf{0}, \mathbf{Q}_{C,k})$, respectively, under Gaussian channel codebooks. 
The purpose of the described message splitting and independent encoding is to incorporate both the edge and cloud decoding approaches in such a way that one signal $\mathbf{s}_{E,k}$ is decoded by the associated AP $i_k$ based on the local uplink signal $\mathbf{y}_{i_k}$, and the other signal $\mathbf{s}_{C,k}$ is decoded by the CS based on quantized signals received from all APs on fronthaul, which will be detailed in this section. 
Transmit covariances $\mathbf{Q}_{E,k}\succeq\mathbf{0}$ and $\mathbf{Q}_{C,k}\succeq\mathbf{0}$ are optimized along with the bit splitting variables $d_{E,k}$ and $d_{C,k}$. Such an optimization policy can achieve performance at least those of edge decoding and cloud decoding schemes.
ID $k$ transmits a superposition of the two signals $\mathbf{s}_{E,k}$ and $\mathbf{s}_{C,k}$, i.e.,
\begin{align}
    \mathbf{x}_k = \mathbf{s}_{E,k} + \mathbf{s}_{C,k}. \label{eq:superposition-coding}
\end{align}
With (\ref{eq:superposition-coding}), the power constraint at ID $k$ can be rewritten as $    \text{tr}\left(\mathbf{Q}_{E,k}\right) + \text{tr}\left(\mathbf{Q}_{C,k}\right) \leq P_{\text{tx}}$.

AP $i$ decodes the edge-decoding signals $\{\mathbf{s}_{E,k}\}_{k\in\mathcal{N}_{I,i}}$ of the assigned IDs $\mathcal{N}_{I,i}$ using its uplink received signal $\mathbf{y}_{i}$. 
We assume independent decoding of those signals to prevent decoding latency. 
Then, the data rate $R_{E,k}$ of each signal $\mathbf{s}_{E,k}$, $k\in\mathcal{N}_{I,i}$, is given by $R_{E,k} = B f_{E,k} (\mathbf{Q})$ \cite{Gamal:11}, where the function $f_{E,k}^{{i_k}} (\mathbf{Q})$ measures the mutual information $I(\mathbf{s}_{E,k}; \mathbf{y}_i)$ and can be computed as
\begin{align}
    f_{E,k}^{{i_k}}(\mathbf{Q}) = \log_2 \det\left( \mathbf{I} + \mathbf{W}_{E,k}^{-1} \mathbf{H}_{{i_k},k} \mathbf{Q}_{E,k} \mathbf{H}_{{i_k},k}^H \right), \label{eq:rate-edge-decoded-signal}
\end{align}
with $\mathbf{Q} \triangleq [[\mathbf{Q}_{E,k}]_{k\in\mathcal{N}_I} \, [\mathbf{Q}_{C,k}]_{k\in\mathcal{N}_I}]$ and
$\mathbf{W}_{E,k} \triangleq \sum\nolimits_{l\in\mathcal{N}_I\setminus\{k\}} \mathbf{H}_{{i_k},l} \mathbf{Q}_{E,l}\mathbf{H}_{{i_k},l}^H + \sum\nolimits_{l\in\mathcal{N}_I} \mathbf{H}_{{i_k},l}\mathbf{Q}_{C,l}\mathbf{H}_{{i_k},l}^H + \sigma_z^2\mathbf{I}$. Here we use the notation $[\mathbf{A}_l]_{l\in\mathcal{N}}$ to denote the matrix obtained by horizontally concatenating the matrices $\mathbf{A}_l$, $l\in\mathcal{N}$.
After the edge decoding, AP $i$ subtracts the impact of the decoded signals $\{\mathbf{s}_{E,k}\}_{k\in\mathcal{N}_{I,i}}$ from the received signal $\mathbf{y}_i$ in \eqref{eq:received-signal-AP-i}. The resulting output $\tilde{\mathbf{y}}_i \triangleq \mathbf{y}_i - \sum\nolimits_{l\in\mathcal{N}_{I,i}} \mathbf{H}_{i,l}\mathbf{s}_{E,l}$ is obtained as 
\begin{align}
    \tilde{\mathbf{y}}_i =  \sum\nolimits_{l\in\mathcal{N}_I \setminus \mathcal{N}_{I,i}} \mathbf{H}_{i,l} \mathbf{s}_{E,l} + \sum\nolimits_{l\in\mathcal{N}_I} \mathbf{H}_{i,l} \mathbf{s}_{C,l} + \mathbf{z}_i. \label{eq:SIC}
\end{align}
Along with the edge-decoded signals $\{\mathbf{s}_{E,k}\}_{k\in\mathcal{N}_{I,i}}$, the signal $\tilde{\mathbf{y}}_i$ is then transferred to the CS to perform the cloud decoding of $\{\mathbf{s}_{C,k}\}_{k\in\mathcal{N}_I}$. 
As the fronthaul link carrying the signal $\tilde{\mathbf{y}}_i$ has a finite capacity of $C_F$ bps, the signal needs to be quantized and compressed to a bit stream prior to fronthauling.

Following the rate-distortion theory \cite{Gamal:11}, the quantized version $\hat{\mathbf{y}}_i$ of $\tilde{\mathbf{y}}_i$, that is obtained by decompressing the bit stream at the CS, can be modeled as
\begin{align}
    \hat{\mathbf{y}}_i = \tilde{\mathbf{y}}_i + \mathbf{q}_i, \label{eq:quantization-model}
\end{align}
where the quantization noise $\mathbf{q}_i$ is independent of $\tilde{\mathbf{y}}_i$ and distributed as $\mathbf{q}_i\sim\mathcal{CN}(\mathbf{0}, \mathbf{\Omega}_i)$.
For a given $\mathbf{\Omega}_i$, the quantized signal $\hat{\mathbf{y}}_{i}$ can be represented by $B \tau_{W} g_i(\mathbf{Q}, \mathbf{\Omega}_i)$ bits \cite{Park:TSIPN21}, which depends on the transmission time on the wireless uplink channel in \eqref{eq:total-latency}.
The function $g_i(\mathbf{Q}, \mathbf{\Omega}_i)$ measuring the number of compression bits per symbol, i.e., the compression rate, is given as \cite{Gamal:11}
\begin{align}
    g_i(\mathbf{Q}, \mathbf{\Omega}_i) = I(\tilde{\mathbf{y}}_i ; \hat{\mathbf{y}}_i) = \log_2 \det \left( \mathbf{I} + \mathbf{\Omega}_i^{-1}\mathbf{W}_{A,i} \right), \label{eq:compression-rate-AP-i}
\end{align}
with $\mathbf{W}_{A,i} \triangleq \sigma_z^2\mathbf{I} + \sum\nolimits_{l\in\mathcal{N}_I\setminus\mathcal{N}_{I,i}} \mathbf{H}_{i,l} \mathbf{Q}_{E,l}\mathbf{H}_{i,l}^H + \sum\nolimits_{l\in\mathcal{N}_I} \mathbf{H}_{i,l}\mathbf{Q}_{C,l}\mathbf{H}_{i,l}^H$.

The CS leverages the fronthaul-received signals $\{\hat{\mathbf{y}}_i\}_{i\in\mathcal{N}_A}$ for decoding the cloud-decoding signals $\{\mathbf{s}_{C,k}\}_{k\in\mathcal{N}_I}$ transmitted by all IDs.
Let $\hat{\mathbf{y}} \triangleq [\hat{\mathbf{y}}_1^H \, \hat{\mathbf{y}}_2^H \, \cdots \hat{\mathbf{y}}_{N_A}^H]^H \in \mathbb{C}^{M_A N_A}$ be a vector stacking all the fronthaul-received signals utilized for cloud decoding. Then, it can be expressed as
\begin{align}
    \hat{\mathbf{y}} = \sum\nolimits_{l\in\mathcal{N}_I} \mathbf{H}_k \mathbf{s}_{C,l} + \sum\nolimits_{l\in\mathcal{N}_I} \mathbf{H}_{E,k} \mathbf{s}_{E,l} + \mathbf{z} + \mathbf{q}, \label{eq:stacked-fronthaul-signals}
\end{align}
where $\mathbf{H}_k \triangleq [\mathbf{H}_{1,k}^H \, \mathbf{H}_{2,k}^H \, \cdots \, \mathbf{H}_{N_A,k}^H]^H$, $\mathbf{z}\triangleq [ \mathbf{z}_1^H \, \mathbf{z}_2^H \, \cdots \, \mathbf{z}_{N_A}^H ]^H$, $\mathbf{q}\triangleq [ \mathbf{q}_1^H \, \mathbf{q}_2^H \, \cdots \, \mathbf{q}_{N_A}^H ]^H$, $\mathbf{H}_{E,k} \triangleq [\mathbf{H}_{E,1,k}^H \, \mathbf{H}_{E,2,k}^H \, \cdots \, \mathbf{H}_{E,N_A,k}^H]^H$, and $\mathbf{H}_{E, i,k} = \mathbf{H}_{i,k}\cdot 1(i\neq i_k)$ with $1(\cdot)$ equal to the indicator function returning 1 if the input statement is true and 0 otherwise. 
The stacked quantization noise signal $\mathbf{q}$ is distributed as $\mathbf{q}\sim \mathcal{CN}(\mathbf{0}, \bar{\boldsymbol{\Omega}})$ with $\bar{\boldsymbol{\Omega}} \triangleq \text{diag}(\{\boldsymbol{\Omega}_i\}_{i\in\mathcal{N}_A})$.

We assume that the CS decodes each signal $\mathbf{s}_{C,k}$ from the fronthaul-received signals $\hat{\mathbf{y}}$ in (\ref{eq:stacked-fronthaul-signals}) after eliminating the impact of the edge-decoding signals $\{\mathbf{s}_{E,l}\}_{l\in\mathcal{N}_I}$, which have been decoded by and forwarded from the APs.
The interference signals $\{\mathbf{s}_{C,l}\}_{l\in\mathcal{N}_I\setminus\{k\}}$ are treated as additive noise to enable time-efficient parallel decoding.
Then, the data rate $R_{C,k}$ of $\mathbf{s}_{C,k}$ is given as $R_{C,k} = B f_{C,k}(\mathbf{Q}, \mathbf{\Omega})$ \cite{Gamal:11},  where
the function $f_{C,k}(\mathbf{Q}, \mathbf{\Omega})$ equals the conditional mutual information $I(\mathbf{s}_{C,k} ; \hat{\mathbf{y}} | \{\mathbf{s}_{E,l}\}_{l\in\mathcal{N}_I})$ with $\boldsymbol{\Omega} \triangleq [\boldsymbol{\Omega}_i]_{i\in\mathcal{N}_A}$ and can be written as
\begin{align}
    f_{C,k}(\mathbf{Q}, \mathbf{\Omega}) = \log_2 \det\left( \mathbf{I} + \mathbf{W}_{C,k}^{-1} \mathbf{H}_k \mathbf{Q}_{C,k} \mathbf{H}_k^H \right). \label{eq:rate-cloud-decoded-signal}
\end{align}
Here, we define 
$\mathbf{W}_{C,k} \triangleq \sigma_z^2\mathbf{I} + \bar{\boldsymbol{\Omega}} + \sum\nolimits_{l\in\mathcal{N}_I\setminus\{k\}}  \mathbf{H}_l\mathbf{Q}_{C,l} \mathbf{H}_l^H$.

The transmission time $\tau_{W}$ from IDs to APs on the wireless uplink channel is given as
\begin{align}
    \tau_{W} = \max_{k\in\mathcal{N}_I, m\in\{E,C\}} \frac{d_{m,k}}{R_{m,k}}. \label{eq:transmission-time-wireless}
\end{align}
Here, the maximum operation is taken over split messages $m\in\{E,C\}$ as well as IDs $k\in\mathcal{N}_{I}$, since the model aggregation at the CS needs the complete reception of both split messages from all IDs.
The transmission time $\tau_{F}$ of the edge-decoded signals and the bit streams representing the quantized signals on the fronthaul links is given as
\begin{align}
    \tau_{F} = \frac{ \max_{i\in\mathcal{N}_A} \Big( \sum\nolimits_{k\in\mathcal{N}_{I,i}} d_{E,k} + B \tau_{W} g_i(\mathbf{Q}, \mathbf{\Omega}_i) \Big) }{ C_F }. \label{eq:transmission-time-frnothaul}
\end{align}

\section{Completion Time Minimization} \label{sec:optimization}

We address the joint optimization of the transmit covariance $\mathbf{Q}$ and the quantization noise covariance $\mathbf{\Omega}$, along with the target local accuracy $\eta_L$ which affects both the numbers $n_{L}$ and $n_{G}$ of local and global iterations, aiming at minimizing the completion time $\tau_{\text{total}}$ in (\ref{eq:total-latency}). 
We assume that the target global accuracy level $\eta_G$ is predetermined, since it is a requirement that should be satisfied by the proposed FL system.
We can formulate the problem as
\begin{subequations} \label{eq:problem-original}
\begin{align}
    \!\!\!\underset{ ^{\mathbf{Q}, \boldsymbol{\Omega}, \boldsymbol{\tau}, \mathbf{d},}_{\mathbf{R}, \eta_L, \mathbf{n}}}{\mathrm{min.}}\,\,\, & n_{G} \left( \tau_C + \tau_{W} + \tau_{F} \right) \label{eq:problem-original-cost} \\
    \mathrm{s.t. }\,\,\,\,\,\, & \tau_C \geq \frac{n_{L} n_C D_k}{ c_k }, \, k\in\mathcal{N}_I, \label{eq:problem-original-latency-computing} \\
    & \tau_{W} \geq \frac{d_{m,k}}{R_{m,k}}, \, k\in\mathcal{N}_I, m\in\{E, C\}, \label{eq:problem-original-latency-wireless} \\
    & \! \tau_{F}\! \geq\! \frac{ \sum\nolimits_{k\in\mathcal{N}_{I,i}}\!\!\! d_{E,k} +\! B\tau_{W} R_{F,i} }{ C_F },  i\in\mathcal{N}_A, \label{eq:problem-original-latency-fronthaul} \\
    & R_{E,k} \leq B f_{E,k}^{{i_k}}(\mathbf{Q}), \, k\in\mathcal{N}_I, \label{eq:problem-original-rate-edge} \\
    & R_{C,k} \leq B f_{C,k}(\mathbf{Q}, \mathbf{\Omega}), \, k\in\mathcal{N}_I, \label{eq:problem-original-rate-cloud} \\
    & R_{F,i} \geq g_i(\mathbf{Q}, \mathbf{\Omega}_i), \, i\in\mathcal{N}_A, \label{eq:problem-original-rate-compression}\\
    & n_{L} \geq -v_L\log_2 \eta_L, \label{eq:problem-original-numIter-local} \\
    & n_{G} \geq \frac{v_G }{ 1 - \eta_L}, \label{eq:problem-original-numIter-global} \\
    & \text{tr}\left(\mathbf{Q}_{E,k}\right) + \text{tr}\left(\mathbf{Q}_{C,k}\right) \leq P_{\text{tx}}, \, k\in\mathcal{N}_I, \label{eq:problem-original-tx-power} \\
    & d_{E,k} + d_{C,k} = d_k, \,\, d_{E,k} \in [0,d_k], \, k\in\mathcal{N}_I, \label{eq:problem-original-split}
\end{align}
\end{subequations}
with the notations $\mathbf{R} \triangleq [[R_{E,k}]_{k\in\mathcal{N}_I} \, [R_{C,k}]_{k\in\mathcal{N}_I} \, [ R_{F,i}]_{i\in\mathcal{N}_A}]$, $\boldsymbol{\tau} \triangleq [\tau_{C} \,\tau_{W} \,\tau_{F}]$, $\mathbf{n} \triangleq [n_{L} \, n_{G}]$, and $\mathbf{d} \triangleq [[d_{E,k}]_{k\in\mathcal{N}_I} \, [d_{C,k}]_{k\in\mathcal{N}_I}]$.
We have also defined the constants $v_G = -(2\beta^2\ln\eta_G)/(\alpha^2\xi)$ and $v_L = 2 / ((2 - \beta\delta)\delta\alpha)$.
The constraints (\ref{eq:problem-original-latency-computing}), (\ref{eq:problem-original-latency-wireless}), and (\ref{eq:problem-original-latency-fronthaul}) impose that the computation time $\tau_C$, the wireless transmission time $\tau_W$, and the fronthaul transmission time $\tau_F$ are lower bounded by the right-hand sides (RHSs) of (\ref{eq:local-computing-time}), (\ref{eq:transmission-time-wireless}), and (\ref{eq:transmission-time-frnothaul}), respectively. 
In order to make the constraint (\ref{eq:transmission-time-frnothaul}) more tractable, we have replaced the function $g_i(\mathbf{Q}, \boldsymbol{\Omega}_i)$ in (\ref{eq:transmission-time-frnothaul}) with a variable $R_{F,i}$ adding a constraint $R_{F,i} \geq g_i(\mathbf{Q}, \boldsymbol{\Omega}_i)$ in (\ref{eq:problem-original-rate-compression}).
In the constraints (\ref{eq:problem-original-rate-edge}) and (\ref{eq:problem-original-rate-cloud}) on the data rates $R_{E,k}$ and $R_{C,k}$ of the edge-decoding and cloud-decoding signals, the RHSs represent the maximum achievable data rates explained in Sec. \ref{sec:rate-splitting-transmission}.
The constraints (\ref{eq:problem-original-numIter-local}) and (\ref{eq:problem-original-numIter-global}) on the numbers of local $n_L$ and global iterations $n_G$ come from (\ref{eq:number-local-iterations}) and (\ref{eq:number-global-iterations}).
The constraint (\ref{eq:problem-original-tx-power}) imposes the transmission power constraints at all IDs, and the last constraint (\ref{eq:problem-original-split}) rewrites (\ref{eq:bit-split-conditions}).

To find an efficient solution of the non-convex problem (\ref{eq:problem-original}), we propose an iterative algorithm, which repeatedly updates the optimization variables by solving the convex problems obtained by convexifying the non-convex functions. First, we replace the non-convex cost function in (\ref{eq:problem-original-cost}) with an epigraph variable $\tau_{\text{total}}$ constrained as
\begin{align}
    \frac{\tau_{\text{total}} }{ n_{G} } \geq \tau_C + \tau_{W} + \tau_{F}. \label{eq:epigraph-total-latency}
\end{align}
From \cite[Prop. 1]{Shen:TN19}, the constraint (\ref{eq:epigraph-total-latency}) holds if there exists $\lambda_{\text{total}}$ satisfying
\begin{align}
    2 \lambda_{\text{total}} \tau_{\text{total}}^{1/2} - \lambda_{\text{total}}^2 n_{G} \geq \tau_C + \tau_{W} + \tau_{F}. \label{eq:convexified-constraint-total-latency}
\end{align}
Note that (\ref{eq:convexified-constraint-total-latency}) is a convex constraint if the auxiliary variable $\lambda_{\text{total}}$ is fixed.
Also, the optimal $\lambda_{\text{total}}$ that makes (\ref{eq:convexified-constraint-total-latency}) equivalent to (\ref{eq:epigraph-total-latency}) is given as
\begin{align}
    \lambda_{\text{total}} = \frac{\tau_{\text{total}}^{1/2} }{ n_{G} }. \label{eq:optimal-lambda-total}
\end{align}
Similarly, we replace the non-convex constraints (\ref{eq:problem-original-latency-wireless}) and (\ref{eq:problem-original-latency-fronthaul}) with the following constraints:
\begin{subequations} \label{eq:convexified-constraint-latency}
\begin{align}
    &2\lambda_{W, m, k} \tau_{W}^{1/2} - \lambda_{W,m,k}^2 d_{m,k} \geq \frac{1}{R_{m,k}},\, k\in\mathcal{N}_I, m\in\{E,C\}, \label{eq:convexified-constraint-latency-wireless} \\
    &\tau_{F} \geq \frac{\sum\nolimits_{k\in\mathcal{N}_I} d_{E,k} + B\mu_{F,i} }{C_F}, \, i\in\mathcal{N}_A, \label{eq:conexified-constraint-latency-fronthaul-1} \\
    & 2\lambda_{F,i} \mu_{F,i}^{1/2} - \lambda_{F,i}^2 \tau_{W} \geq R_{F,i}, \, i\in\mathcal{N}_A, \label{eq:convexified-constraint-latency-fronthaul-2}
\end{align}
\end{subequations}
which become equivalent to (\ref{eq:problem-original-latency-wireless}) and (\ref{eq:problem-original-latency-fronthaul}) if
\begin{align}
    \lambda_{W,m,k} = \frac{\tau_{W}^{1/2} }{ d_{m,k} } \,\,\, \text{and} \,\,\,
    \lambda_{F,i} = \frac{\mu_{F,i}^{1/2} }{ \tau_{W} }.  \label{eq:optimal-lambda-latency}
\end{align}

According to \cite[Thm. 2]{Shen:TN19}, the following constraint is a sufficient condition for the non-convex constraint (\ref{eq:problem-original-rate-edge}):
\begin{align}
    \frac{R_{E,k}}{B} &\leq \log_2\det\left(\mathbf{I} + \boldsymbol{\Gamma}_{E,k}\right) + \frac{1}{\ln 2}\text{tr}(\boldsymbol{\Gamma}_{E,k}) \nonumber \\
    & + \frac{1}{\ln 2}\, \text{tr}\Big( \left(\mathbf{I} + \boldsymbol{\Gamma}_{E,k}\right) \Big( 2\tilde{\mathbf{Q}}_{E,k}^H \mathbf{H}_{i_k, k}^H \boldsymbol{\Theta}_{E,k} - \boldsymbol{\Theta}_{E,k}^H \left( \mathbf{W}_{E,k} + \mathbf{H}_{i_k,k} \mathbf{Q}_{E,k} \mathbf{H}_{i_k,k}^H \right) \boldsymbol{\Theta}_{E,k} \Big)  \Big), \label{eq:convexified-constraint-rate-edge}
\end{align}
for $k\in\mathcal{N}_A$, 
with $\tilde{\mathbf{Q}}_{E,k} = \mathbf{Q}_{E,k}^{1/2}$.
The constraint (\ref{eq:convexified-constraint-rate-edge}) is convex for fixed $\boldsymbol{\Gamma}_{E,k}$ and $\boldsymbol{\Theta}_{E,k}$ and is equivalent to (\ref{eq:problem-original-rate-edge}) if
\begin{subequations} \label{eq:optimal-Gamma-Theta-edge}
\begin{align}
    &\boldsymbol{\Gamma}_{E,k} = \tilde{\mathbf{Q}}_{E,k}^H \mathbf{H}_{i_k,k}^H \mathbf{W}_{E,k}^{-1} \mathbf{H}_{i_k, k}\tilde{\mathbf{Q}}_{E,k}, \label{eq:optimal-Gamma-edge} \\
    & \mathbf{\Theta}_{E,k} = \left( \mathbf{W}_{E,k} + \mathbf{H}_{i_k, k} \mathbf{Q}_{E,k} \mathbf{H}_{i_k,k}^H \right)^{-1} \mathbf{H}_{i_k, k} \tilde{\mathbf{Q}}_{E,k}. \label{eq:optimal-Theta-edge}
\end{align}
\end{subequations}
Likewise, defining $\tilde{\mathbf{Q}}_{C,k} = \mathbf{Q}_{C,k}^{1/2}$, we replace the non-convex constraint (\ref{eq:problem-original-rate-cloud}) with the following constraint:
\begin{align}
    \frac{R_{C,k}}{B} & \leq \log_2\det\left( \mathbf{I} + \boldsymbol{\Gamma}_{C,k} \right) - \frac{1}{\ln 2}\text{tr} \left( \boldsymbol{\Gamma}_{C,k} \right) \nonumber \\
    & + \frac{1}{\ln 2} \text{tr} \Big( \left( \mathbf{I} + \boldsymbol{\Gamma}_{C,k} \right) \Big( 2\tilde{\mathbf{Q}}_{C,k}^H \mathbf{H}_k^H \mathbf{\Theta}_{C,k} -  \mathbf{\Theta}_{C,k}^H \left( \mathbf{W}_{C,k} + \mathbf{H}_k \mathbf{Q}_{C,k} \mathbf{H}_k^H \right)\mathbf{\Theta}_{C,k} \Big)\Big), \label{eq:convexified-constraint-rate-cloud}
\end{align}
for $k\in\mathcal{N}_A$, which becomes equivalent to (\ref{eq:problem-original-rate-cloud}) when
\begin{subequations} \label{eq:optimal-Gamma-Theta-cloud}
\begin{align}
    &\boldsymbol{\Gamma}_{C,k} = \tilde{\mathbf{Q}}_{C,k}^H \mathbf{H}_k^H \mathbf{W}_{C,k}^{-1} \mathbf{H}_k \tilde{\mathbf{Q}}_{C,k}, \label{eq:optimal-Gamma-cloud} \\
    &\mathbf{\Theta}_{C,k} = \left( \mathbf{W}_{C,k} + \mathbf{H}_k \mathbf{Q}_{C,k} \mathbf{H}_k^H \right)^{-1} \mathbf{H}_k \tilde{\mathbf{Q}}_{C,k}.\label{eq:optimal-Theta-cloud}
\end{align}
\end{subequations}
In addition, Lemma 1 in \cite{Zhou:TSP16} states that the non-convex constraint in (\ref{eq:problem-original-rate-compression}) holds if the following condition is satisfied:
\begin{align}
    R_{F,i} & \geq \log_2\det\left( \boldsymbol{\Sigma}_{F,i} \right) + \frac{\text{tr}\left( \boldsymbol{\Sigma}_{F,i}^{-1}\left( \mathbf{W}_{A,i} + \boldsymbol{\Omega}_i \right)  \right) }{ \ln 2 }  - \frac{M_A}{\ln 2} - \log_2\det\left(\boldsymbol{\Omega}_i\right), \, i\in\mathcal{N}_A. \label{eq:convexified-constraint-rate-compression}
\end{align}
The two constraints (\ref{eq:problem-original-rate-compression}) and (\ref{eq:convexified-constraint-rate-compression}) become equivalent if
\begin{align}
    \boldsymbol{\Sigma}_{F,i} = \mathbf{W}_{A,i} + \boldsymbol{\Omega}_i. \label{eq:optimal-Sigma}
\end{align}

Using (\ref{eq:convexified-constraint-total-latency}), (\ref{eq:convexified-constraint-latency}), (\ref{eq:convexified-constraint-rate-edge}), (\ref{eq:convexified-constraint-rate-cloud}), and (\ref{eq:convexified-constraint-rate-compression}) and 
including $\boldsymbol{\lambda}\triangleq[\lambda_{\text{total}} \, [\lambda_{W,m,k}]_{m\in\{E,C\}, k\in\mathcal{N}_I} \, [\lambda_{F,i}]_{i\in\mathcal{N}_A}] $, $\boldsymbol{\Gamma} \triangleq [\boldsymbol{\Gamma}_{m,k}]_{m\in\{E,C\}, k\in\mathcal{N}_I}$, $\boldsymbol{\Theta} \triangleq [\boldsymbol{\Theta}_{m,k}]_{m\in\{E,C\}, k\in\mathcal{N}_I}$, and $\boldsymbol{\Sigma} \triangleq [\boldsymbol{\Sigma}_{F,i}]_{i\in\mathcal{N}_A}$ into the set of optimization variables,
we can equivalently restate problem (\ref{eq:problem-original}) as
\begin{align}
    \underset{ ^{\tilde{\mathbf{Q}}, \boldsymbol{\Omega}, \boldsymbol{\tau}, \mathbf{d}, \mathbf{R}, }_{\eta_L, \mathbf{n},\boldsymbol{\lambda}, \boldsymbol{\Gamma}, \boldsymbol{\Theta}, \boldsymbol{\Sigma}}  }{\mathrm{min.}}\,\,\, & \tau_{\text{total}} \label{eq:problem-modified} \\
    \mathrm{s.t. }\,\,\,\,\,\, &  \text{(\ref{eq:convexified-constraint-total-latency}), (\ref{eq:convexified-constraint-latency}), (\ref{eq:convexified-constraint-rate-edge}), (\ref{eq:convexified-constraint-rate-cloud}), (\ref{eq:convexified-constraint-rate-compression}),} \nonumber \\
    & \text{(\ref{eq:problem-original-latency-computing}), (\ref{eq:problem-original-numIter-local}), (\ref{eq:problem-original-numIter-global}), (\ref{eq:problem-original-split}),} \nonumber \\
    & \sum\nolimits_{m\in\{E,C\}}||\tilde{\mathbf{Q}}_{m,k}||_F^2\leq P_{\text{tx}}, \, k\in\mathcal{N}_I. \nonumber
\end{align}
Note that problem (\ref{eq:problem-modified}) is tackled with respect to $\tilde{\mathbf{Q}} \triangleq [ [\tilde{\mathbf{Q}}_{E,k}]_{k\in\mathcal{N}_I} [\tilde{\mathbf{Q}}_{C,k}]_{k\in\mathcal{N}_I} ]$ instead of $\mathbf{Q}$.
Although the problem (\ref{eq:problem-modified}) is still non-convex, we obtain a convex problem if the additional variables $\{\boldsymbol{\lambda}, \boldsymbol{\Gamma}, \boldsymbol{\Theta}, \boldsymbol{\Sigma}\}$ are fixed.
Also, the optimal $\{\boldsymbol{\lambda}, \boldsymbol{\Gamma}, \boldsymbol{\Theta}, \boldsymbol{\Sigma}\}$ for fixed other variables can be computed with the closed-form expressions in (\ref{eq:optimal-lambda-total}), (\ref{eq:optimal-lambda-latency}), (\ref{eq:optimal-Gamma-Theta-edge}), (\ref{eq:optimal-Gamma-Theta-cloud}), and (\ref{eq:optimal-Sigma}).
Therefore, with a proper choice of initial variables $\{\tilde{\mathbf{Q}},\boldsymbol{\Omega}, \mathbf{d}, \eta_L, \mathbf{n}\}$, a sequence of monotonically decreasing cost values can be obtained by alternately updating the variable sets $\{\tilde{\mathbf{Q}},\boldsymbol{\Omega}, \mathbf{d}, \eta_L, \mathbf{n}\}$ and $\{\boldsymbol{\lambda}, \boldsymbol{\Gamma}, \boldsymbol{\Theta}, \boldsymbol{\Sigma}\}$.
We summarize the detailed algorithm in Algorithm 1. 
Because the optimal completion time $\tau_{\text{total}}$ is not divergent, the solution of Algorithm 1 converges to a locally optimal point\footnote{{We leave the development of an efficient algorithm that finds a global optimal solution of problem (\ref{eq:problem-original}) as a future work.}} of the problem (\ref{eq:problem-modified}).
In Step 6, $e_{\text{th}}$ indicates the threshold value for judging the convergence in terms of cost function value, and $t_{\max}$ is the maximum allowed number of iterations.
After convergence, each transmit covariance matrix $\mathbf{Q}_{m,k}$ is obtained as $\mathbf{Q}_{m,k}\leftarrow \tilde{\mathbf{Q}}_{m,k}\tilde{\mathbf{Q}}_{m,k}^H$ for $m\in\{E,C\}$, $k\in\mathcal{N}_I$.

\begin{algorithm}
\caption{Proposed alternating optimization algorithm}

\textbf{1.} Initialize $\{\tilde{\mathbf{Q}},\boldsymbol{\Omega}\}$ as arbitrary matrices that satisfy the constraints (\ref{eq:problem-original-tx-power}), and set $t\leftarrow 1$.

\textbf{2.} Compute the completion time in (\ref{eq:total-latency}) with the initial $\{\tilde{\mathbf{Q}},\boldsymbol{\Omega}\}$ and store to $\tau_{\text{total}}^{(t)}$.

\textbf{3.} Set the variables $\{\boldsymbol{\lambda}, \boldsymbol{\Gamma}, \boldsymbol{\Theta}, \boldsymbol{\Sigma}\}$ as (\ref{eq:optimal-lambda-total}), (\ref{eq:optimal-lambda-latency}), (\ref{eq:optimal-Gamma-Theta-edge}), (\ref{eq:optimal-Gamma-Theta-cloud}), and (\ref{eq:optimal-Sigma}).

\textbf{4.} Update the primary variables $\{\tilde{\mathbf{Q}},\boldsymbol{\Omega}, \mathbf{d}, \eta_L\}$ as a solution of the convex problem obtained by fixing $\{\boldsymbol{\lambda}, \boldsymbol{\Gamma}, \boldsymbol{\Theta}, \boldsymbol{\Sigma}\}$ in problem (\ref{eq:problem-modified}).

\textbf{5.} Compute the completion time in (\ref{eq:total-latency}) with the updated $\{\tilde{\mathbf{Q}},\boldsymbol{\Omega}\}$ and store to $\tau_{\text{total}}^{(t+1)}$.

\textbf{6.} Stop if $|\tau_{\text{total}}^{(t+1)} - \tau_{\text{total}}^{(t)}| < e_{\text{th}}$ or $t \geq t_{\max}$. Otherwise, go back to Step 3 with $t\leftarrow t + 1$.

\end{algorithm}

The complexity of Algorithm 1 is the product of the number of iterations and the complexity of each iteration, i.e., Steps 3, 4, and 5. 
The latter is dominated by the complexity required to solve the convex problem at Step 4, i.e., the convex problem obtained by fixing $\{\boldsymbol{\Lambda}, \boldsymbol{\Gamma}, \boldsymbol{\Theta}, \boldsymbol{\Sigma}\}$ in problem (\ref{eq:problem-modified}).
It was shown in \cite[p. 4]{BenTal:19} that the worst-case complexity of finding a solution of a generic convex optimization problem is given as $\mathcal{O}( n_v(n_v^3+n_c) \log(1/\epsilon) )$. Here $n_v$ indicates the number of optimization variables, $n_c$ denotes the number of arithmetic operations required to compute the objective and constraint functions, and $\epsilon$ represents the desired error tolerance level.
The numbers $n_v$ and $n_c$ for the convex problem of Step 4 are given as $n_v=\mathcal{O}( \max\{M_I^2 N_I, M_A^2 N_A\} )$ and $n_c=\mathcal{O}( M_A^2 N_A^2 M_I^2 N_I )$, respectively.
The number of iterations required for convergence is determined by the convergence rate, but its analysis for general successive convex approximation algorithms is still an open problem and is left as a future work.

\section{Numerical Results}

We evaluate the performance of the fog-RAN-assisted FL system which adopts the proposed optimized rate-splitting scheme and compare with the following schemes: \textit{Edge-decoding only} ($d_{E,k}=d_k$, $\forall k\in\mathcal{N}_I$) and \textit{cloud-decoding only} schemes ($d_{E,k}=0$, $\forall k\in\mathcal{N}_I$). IDs and APs are randomly distributed within an area of radius 200 m while satisfying the minimum separation of 10 m.
The Rayleigh fading model is adopted as $\mathbf{H}_{i,k} = \rho_0 ( \text{dist}_{i,k}/ \text{dist}_0 )^{-\gamma/2} \tilde{\mathbf{H}}_{i,k}$ where $\text{dist}_{i,k}$ is the distance between ID $k$ and AP $i$, $\rho_0=10$ dB stands for the path-loss at reference distance $\text{dist}_0 = 50$ m, and the small-scale fading $\tilde{\mathbf{H}}_{i,k}$ is distributed as $\text{vec}(\tilde{\mathbf{H}}_{i,k}) \sim \mathcal{CN}(\mathbf{0}, \mathbf{I})$.
The parameters of local computing tasks are set to $D_k=500$, $d_k=28\cdot 10^3$, $c_k = 3\cdot 10^9$, and $n_C \in \{200, 2000\}$ \cite{Vu:TWC20, Yang:TWC21, Yao:TCCN}.
We also fix $P_{\text{tx}}/\sigma_z^2 = 10$ dB, $B=20$ MHz, $({M_I, M_A}) = (1,2)$, and $\eta_G = 10^{-3}$ unless stated otherwise.

\begin{figure}
\centering\includegraphics[width=0.80\linewidth]{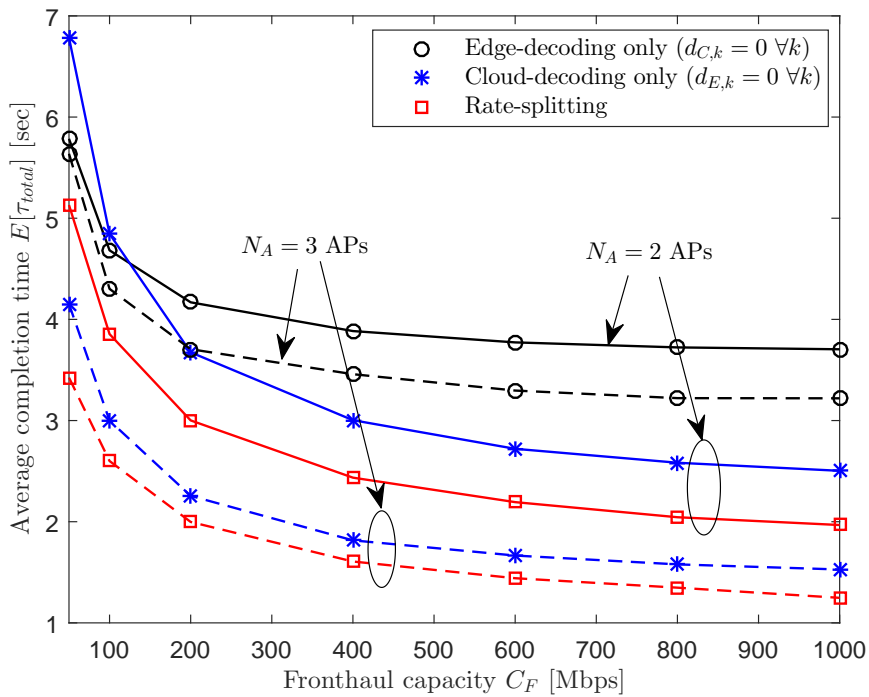}
\vspace{-4mm}
\caption{{\label{fig:latency-vs-CF}Average completion time $\mathbb{E}[\tau_{\text{total}}]$ versus the fronthaul capacity $C_F$}}
\end{figure}

Fig. \ref{fig:latency-vs-CF} depicts the average completion time $\mathbb{E}[\tau_{\text{total}}]$ versus the fronthaul capacity $C_F$ for $N_I=10$, $N_A\in\{2,3\}$, and $n_C = 200$. 
The performance of all schemes is improved with a larger $C_F$ due to the savings in the fronthaul latency $\tau_{F}$.
The cloud-decoding scheme gets more benefit than the edge-decoding scheme thanks to more pronounced impact of cooperative cloud detection.
For all simulated scenarios, the rate-splitting transmission with the proposed optimization algorithm provides relevant gains compared to the baseline schemes. 
The gain over the cloud-decoding scheme is more significant for the case with less APs, in which the impact of cooperative cloud decoding would be marginal.

\begin{figure}
\centering\includegraphics[width=0.80\linewidth]{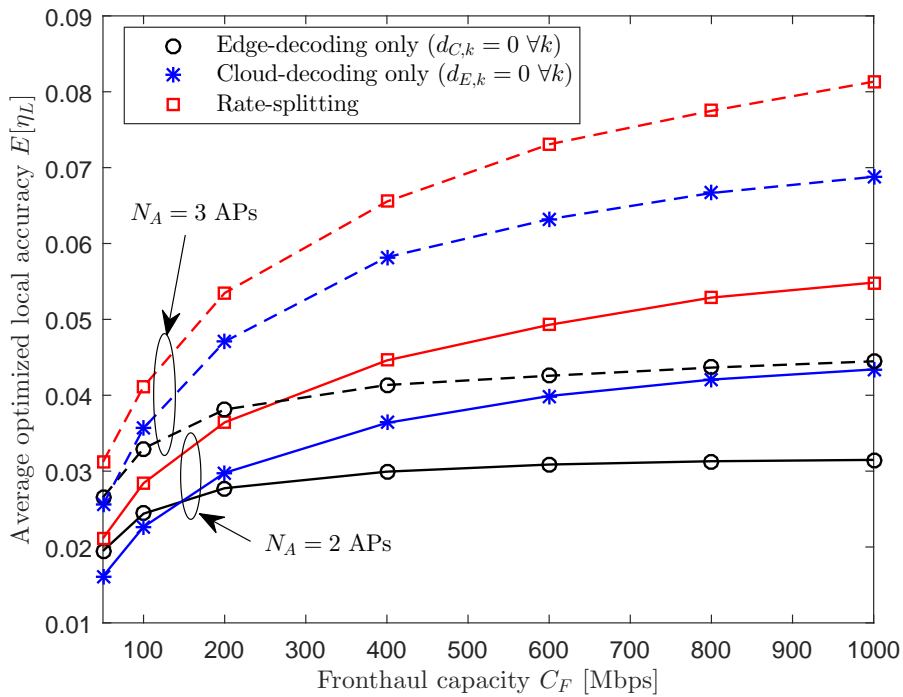}
\vspace{-4mm}
\caption{{\label{fig:localAccuracy-vs-CF}Average optimized local accuracy level  $\mathbb{E}[\eta_L]$ versus $C_F$}}
\end{figure}

Fig. \ref{fig:localAccuracy-vs-CF} plots the average optimized local accuracy level $\mathbb{E}[\eta_L]$ with respect to the fronthaul capacity $C_F$ for the same setup in Fig. \ref{fig:latency-vs-CF}. 
As $C_F$ increases, the impact of the fronthaul latency $\tau_{F}$ on the completion time (\ref{eq:total-latency}) decreases. 
Thus, it is desirable to choose a larger local accuracy level $\eta_L$ to suppress the computation time $\tau_C$. 
In a similar vein, the optimized local accuracy of the rate-splitting scheme is larger than those of the baseline schemes, since it requires shorter transmission times $\tau_{W}$ and $\tau_{F}$, which lead to a larger dominance of the computation time $\tau_C$ than the other schemes.

\begin{figure}
\centering\includegraphics[width=0.80\linewidth]{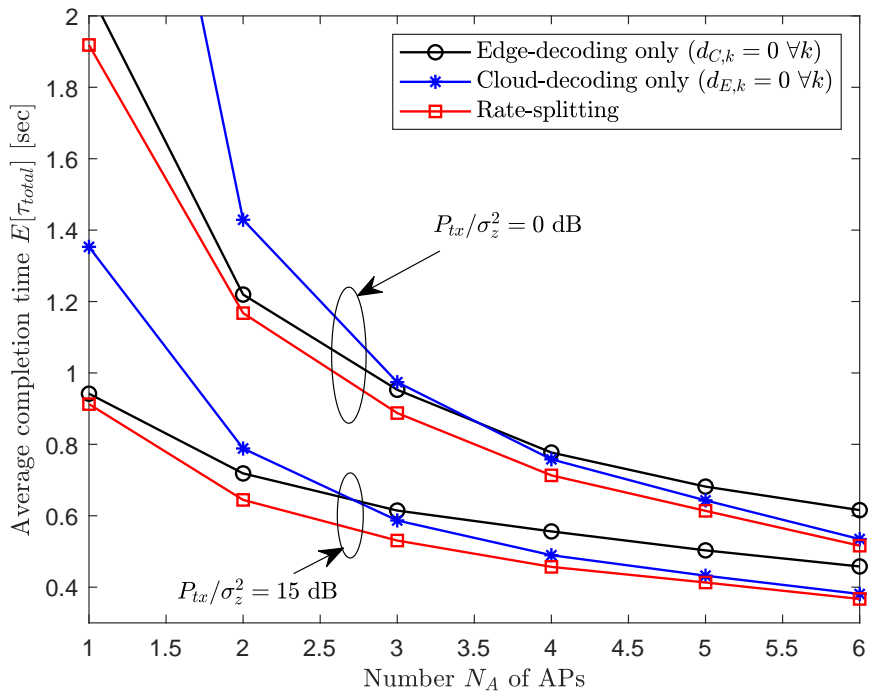}
\vspace{-4mm}
\caption{{\label{fig:latency-vs-NA}Average completion time $\mathbb{E}[\tau_{\text{total}}]$ versus the number $N_A$ of APs}}
\end{figure}

In Fig. \ref{fig:latency-vs-NA}, we plot the average completion time $\mathbb{E}[\tau_{\text{total}}]$ versus the number $N_A$ of APs for $N_I=7$, $(M_I,M_A)=(2,2)$, $n_C=200$, $C_F=500$ Mbps, $B=100$ MHz, and $P_{\text{tx}}/\sigma_z^2\in\{0, 15\}$ dB.
The cloud-decoding strategy outperforms the edge-decoding scheme when there are many APs due to more effective interference management capability enabled by joint decoding at CS.
However, if only a few APs are involved, the edge-decoding scheme shows lower latency, which means that the impact of quantization noise signals of the cloud-decoding scheme cannot be counter-balanced by the advantage of cooperative decoding among small number of APs.

\section{Conclusion}

We have studied a fog-RAN-assisted FL system with finite-capacity fronthaul links. To enable flexible hybrid edge and cloud decoding strategies, we have proposed a rate-splitting transmission scheme and have tackled the problem of minimizing the completion time of FL with a given target global accuracy requirement. 
Via numerical results, we have checked the performance gain of the proposed optimized rate-splitting scheme over benchmark schemes and observed the impacts of various system parameters on the performance gains and the optimized local accuracy level.
Among interesting open problems, we mention the optimization of global accuracy level and identifying its relation to test accuracy in actual learning tasks.

\end{document}